\documentclass[12pt]{article}
\usepackage{amsmath,amssymb}
\usepackage{graphicx,psfrag,epsf}
\usepackage{enumerate}
\usepackage{natbib}
\usepackage{url} 

\newcommand{\blind}{0}

\begin{document}

\def\spacingset#1{\renewcommand{\baselinestretch}%
{#1}\small\normalsize} \spacingset{1}


\if0\blind
{
  \title{\bf Coefficients of Determination for Mixed-Effects Models}
  \author{Dabao Zhang
    \hspace{.2cm}\\
    Department of Statistics, Purdue University}
  \maketitle
} \fi

\if1\blind
{
  \bigskip
  \bigskip
  \bigskip
  \begin{center}
    {\LARGE\bf Coefficients of Determination\\ \vskip12pt for Mixed-Effects Models}
\end{center}
  \medskip
} \fi

\bigskip
\begin{abstract}
The coefficient of determination is well defined for linear models and its extension is long wanted for mixed-effects models in agricultural, biological, and ecological research. We revisit its extension to define measures for proportions of variation explained by the whole model, fixed effects only, and random effects only. We propose to calculate unexplained variations conditional on individual random and/or fixed effects so as to keep individual heterogeneity brought by available predictors. While naturally defined for linear mixed models, these measures can be defined for a generalized linear mixed model using a distance measured along its variance function, accounting for its heteroscedasticity. We demonstrate the promising performance and utility of our proposed methods via simulation studies as well as applications to real data sets in agricultural and ecological studies.
\end{abstract}

\noindent%
{\it Keywords:}  Exponential family distribution; Generalized linear mixed model; Linear mixed model; Quasi-model; $R^2$; Variance function
\vfill

\newpage
\spacingset{1.45} 


\section{Introduction} \label{sec:intro}

Mixed-effect models are widely used in agricultural, biological, and ecological research to understand the variation components of a response variable \citep{Gbur2012, Zuur2009}. The law of total variance provides a theoretical basis for defining the coefficient of determination, also known as $R^2$, for linear models and sheds light on defining similar measures for mixed-effects models.



Suppose that a linear mixed model \citep{McCulloch2008} is considered for observed response variable $Y_{ij}$ from the $j$-th individual inside the $i$-th cluster, with $j=1, \cdots, n_i$ and $i=1, \cdots, m$. In general, we write the corresponding linear mixed model as,
\begin{eqnarray} \label{eqn-mixedmodel}
Y_{ij} = \eta^F_{ij} + \eta^R_{ij} + \epsilon_{ij}, \ \ \ \epsilon_{ij}\sim N(0, \sigma^2),
\end{eqnarray}
where $\eta^F_{ij}$ and $\eta^R_{ij}$ respectively summarize all fixed and random effects on the response variable with
\[
\eta^R_{ij}\mid \tau^2_{ij}\sim N(0,\tau^2_{ij}).
\]
Note that $\{\epsilon_{ij}, j=, \cdots, m\}$ may be an autocorrelated series in, for example, longitudinal studies \citep{Larid1982}. For construction simplicity, we will not emphasize such correlation as it does not affect the unbiasedness of estimated variances in the below although the estimation may not be optimal.

Similar to linear models, the law of total variance provides a clear path to extend $R^2$ for linear mixed models, measuring the proportion of variation in the dependent variable modeled by fixed effects, random effects, or both \citep{Xu2003,Nakagawa2013,Nakagawa2017,Jaeger2017,Jaeger2019}. However, unlike usual calculation that heavily relies on estimated error variance $\sigma^2$, we instead propose to calculate unexplained variations conditional on individual random and/or fixed effects so as to keep individual heterogeneity brought by available predictors which also shed lights on their extensions to more general models.

The inherent heteroscedasticity makes it difficult to properly define $R^2$ for generalized linear mixed models. Indeed, such heteroscedasticity also challenges the proper definition of $R^2$ for generalized linear models \citep{McCullagh1989}. Therefore, many different measures have been proposed to define $R^2$ for generalized linear models from different aspects of view \citep{Cameron1997,Cox1989,Maddala1983,Magee1990,Nagelkerke1991,Zhang2017}. However, it is difficult to extend these measures to account for random effects involved in generalized linear mixed models.

A common strategy to define $R^2$ for generalized linear mixed models, adopted by \citet{Nakagawa2013} and \citet{Nakagawa2017}, is to recognize the linear function presented by the link function $g(\cdot)$, i.e.,
\begin{eqnarray} \label{eqn-glmm}
g(E[Y_{ij}\mid \eta^F_{ij},\eta^R_{ij}]) = \eta^F_{ij} + \eta^R_{ij},
\end{eqnarray}
and instead construct $R^2$ for a transformed linear mixed model for $g(Y_{ij})$. However, such measures rely on the specified link function and even the approximation method used to calculate the error variance \citep{Nakagawa2017}. On the other hand, the link function does not necessarily provide homoscedastic variance on the error term, which is the primary challenge in extending classical $R^2$ from linear models to generalized linear models, although it may describe a linear relationship of all effects on $g(E[Y_{ij}\mid \eta^F_{ij},\eta^R_{ij}])$, and present additive variance components in $g(Y_{ij})$.

A critical concern of defining $R^2$ based on a transformed linear model is that proportions of different variance components in $g(Y_{ij})$ may not represent the genuine proportions of different variance components in $Y_{ij}$. For example, it is well-known that the latent linear model of a probit model may present a much higher $R^2$, but the binomial response still holds a lot of uncertainty, which is well recognized in the study of genetic heritability, see \citet{Dempster1950}.

Recently \citet{Zhang2017} showed that quantifying the variation change along the variance function can measure explained variation of a heteroscedastic response variable and hence proposed to define a variance-function-based $R^2$. Unlike other likelihood-based measures, such a measure neither overstates the proportion of explained variation, nor demands the specification of likelihood functions. While it only requires specification of the link function and variance function, it reduces to classical $R^2$ for generalized linear models so it is conceptually consistent with classical $R^2$. We thus use the variance-function-based measures to define $R^2$ for generalized linear mixed models.


In the next section, we revisit the definition of coefficients of determination to account for explained variation by fixed effects, random effects, or both in linear mixed models, and propose our calculation emphasizing individual heterogeneity. In Section 3, we extend these measures for generalized linear mixed models by using the variance-function-based distance \citep{Zhang2017}. Our simulation studies to compare different measures are shown in Section 4. We also compare these measures in Section 5 by applying them to three sets of real data from agricultural and ecological studies, and conclude with a discussion in Section 6.

\section{Linear Mixed Models}


For the linear mixed model~(\ref{eqn-mixedmodel}), we can follow the law of total variance and define the proportion of variation in $Y_{ij}$ modeled by the fixed effects as
\begin{eqnarray} \label{eqn-lmmr2f}
\rho_F^2 = 1-\frac{E[\mbox{var}(Y_{ij}\mid \eta^F_{ij})]}{\mbox{var}(Y_{ij})} =
 1-\frac{E[(Y_{ij}-E[Y_{ij}\mid \eta^F_{ij}])^2]}{E[(Y_{ij}-E[Y_{ij}])^2]}.
\end{eqnarray}

With $\eta^F_{ij}$ estimated by $\hat{\eta}^F_{ij}$ and $E[Y_{ij}]$ estimated by the sample average $\bar{Y}_{\cdot\cdot}$, we have the following estimate of $\rho_F^2$,
\begin{eqnarray} \label{eqn-lmmr2fest}
R_F^2 = 1-\frac{\sum_{i,j}(Y_{ij}-\hat{\eta}^F_{ij})^2}{\sum_{i,j}(Y_{ij}-\bar{Y}_{\cdot\cdot})^2}.
\end{eqnarray}
Since
\[
E[(Y_{ij}-E[Y_{ij}\mid \eta^F_{ij}])^2] = E[\tau^2_{ij}]+\sigma^2,
\]
we may also take estimated variance components to construct $R_F^2$ to estimate $\rho_F^2$, see, e.g., \citet{Xu2003}, \citet{Nakagawa2013}, \citet{Nakagawa2017}, and \citet{Jaeger2017}. As our revisit to defining $R^2$ in linear mixed models also serves to shed light on their extensions to generalized linear mixed models \citep{McCullagh1989}, we will not pursue this avenue as it cannot manage the heterogeneity in generalized linear mixed models.

The proportion of variation in $Y_{ij}$ modeled by both fixed and random effects can be similarly defined as
\begin{eqnarray} \label{eqn-lmmr2fr}
\rho_M^2 = 1-\frac{E[\mbox{var}(Y_{ij}\mid \eta^F_{ij}, \eta^R_{ij})]}{\mbox{var}(Y_{ij})} = 1-\frac{E[(Y_{ij}-E[Y_{ij}\mid \eta^F_{ij}, \eta^R_{ij}])^2]}{E[(Y_{ij}-E[Y_{ij}])^2]}.
\end{eqnarray}
With $\mbox{var}(Y_{ij}\mid \eta^F_{ij}, \eta^R_{ij})=\sigma^2$, it is tempting to estimate $\rho_M^2$ on the basis of $\rho_M^2=1-\sigma^2/\mbox{var}(Y_{ij})$ as in \citet{Xu2003}, \citet{Nakagawa2013}, \citet{Nakagawa2017}, \citet{Jaeger2017}, and \citet{Ives2019}. However, the total variation in the response variable is described by $SST=\sum_{i,j} (Y_{ij}-\bar{Y}_{\cdot\cdot})^2$, and we thus would rather to calculate the total unexplained variation by emphasizing individual heterogeneity of $\tau_{ij}$ and the contribution of individual observation, which also help the extension to generalized linear models. 

Note that,
\[
E\left[(Y_{ij}-E[Y_{ij}\mid \eta^F_{ij}, \eta^R_{ij}])^2\right] = E\left[E[(Y_{ij}-E[Y_{ij}\mid \eta^F_{ij}, \eta^R_{ij}])^2\mid Y_{ij}, \eta^F_{ij}, \tau_{ij}^2, \sigma^2]\right],
\]
implying that each observation contributes $E\left[(Y_{ij}-E[Y_{ij}\mid \eta^F_{ij}, \eta^R_{ij}])^2\mid Y_{ij}, \eta^F_{ij}, \tau_{ij}^2, \sigma^2\right]$ with observed value $Y_{ij}$ and estimable parameters in $\eta^F_{ij}, \tau^2_{ij}$, and $\sigma^2$. That is, the expectation is on the random variable $\eta^R_{ij}$ conditional on the observed values and these estimable parameters. With the conditional distribution
\[
\eta^R_{ij}\mid Y_{ij}, \eta^F_{ij}, \tau^2_{ij}, \sigma^2 \sim N\left(\frac{\tau^2_{ij}}{\sigma^2+\tau^2_{ij}}(Y_{ij}-\eta^F_{ij}), \frac{\sigma^2\tau^2_{ij}}{\sigma^2+\tau^2_{ij}}\right),
\]
we have
\[
E[(Y_{ij}-E[Y_{ij}\mid \eta^F_{ij}, \eta^R_{ij}])^2\mid Y_{ij}, \eta^F_{ij}, \tau_{ij}^2, \sigma^2] = \left(\frac{\sigma^2}{\sigma^2+\tau^2_{ij}}\right)^2 \left(Y_{ij}-\eta^F_{ij}\right)^2 + \frac{\sigma^2\tau^2_{ij}}{\sigma^2+\tau^2_{ij}}.
\]
Therefore, $\rho_M^2$ will be estimated by
\begin{eqnarray} \label{eqn-lmmr2frest}
R_M^2 = 1-\frac{
\sum_{i,j} \frac{\hat{\sigma}^2}{\hat{\sigma}^2+\hat{\tau}^2_{ij}}
\left\{\hat{\tau}_{ij}^2 +  \frac{\hat{\sigma}^2}{\hat{\sigma}^2+\hat{\tau}^2_{ij}}\left(Y_{ij}-\hat{\eta}^F_{ij}\right)^2\right\}
}{\sum_{i,j}(Y_{ij}-\bar{Y}_{\cdot\cdot})^2}.
\end{eqnarray}



The proportion of variation in $Y_{ij}$ modeled by random effects can be simply defined as
\begin{eqnarray} \label{eqn-lmmr2r}
\rho_{R}^2 = \rho_M^2 - \rho_{F}^2 = \frac{E[(Y_{ij}-E[Y_{ij}\mid \eta^F_{ij}, \tau^2_{ij}])^2] - E[(Y_{ij}-E[Y_{ij}\mid \eta^F_{ij}, \eta^R_{ij}])^2]}{\mbox{var}(Y_{ij})},
\end{eqnarray}
and can be estimated as
\begin{eqnarray} \label{eqn-lmmr2rest}
R_R^2 = R_M^2 - R_F^2.
\end{eqnarray}

Such a simple definition on $\rho_R^2$ assumes that $\rho_F^2$ has a priority over $\rho_R^2$, that is, when fixed and random effects overlapped, we would rather know the proportion of variation explained by available fixed effects, and therefore use $\rho_R^2$ to measure the proportion of variable additionally explained by random effects. 

\section{Generalized Linear Mixed Models}

For the generalized linear mixed model (\ref{eqn-glmm}), the variance of $Y_{ij}$, given both fixed and random effects, can be specified via a dispersion parameter $\phi$ and a known variance function $V(\cdot)$, i.e.,
\[
\mbox{var}(Y_{ij}\mid \eta^F_{ij}, \eta^R_{ij}) = \phi V(g^{-1}(\eta^F_{ij}+\eta^R_{ij})).
\]
In general, as long as the mean $g^{-1}(\eta^F_{ij}+\eta^R_{ij})$ can be modeled well and linked appropriately to a set of predictors, a generalized linear model with known variance function $V(\cdot)$ can be investigated for the utility of the involved predictors.

The variance function describes the effect of the mean on the variation of the response variable besides the dispersion parameter. For a response variable with its mean moving from $a$ to $b$, its variation changes accordingly along the variance function from $\phi V(a)$ to $\phi V(b)$. Taking advantage of relationship between mean and variance defined by the variance function, we measure the variation change along the variance function \citep{Zhang2017}. That is, the variation change of the response variable should be measured using, instead of $(a-b)^2$, the squared length of the variance function $V(\cdot)$ between $V(a)$ to $V(b)$,
\[
d_V(a, b) = \left\{\int_a^b \sqrt{1+[V'(t)]^2} dt \right\}^2.
\]
As show in Table~\ref{Table-dv}, $d_V(\cdot,\cdot)$ may dramatically differ than the Euclidean distance when the variance function is nonlinear.

\begin{center}
\begin{table}[!ht]
\caption{\label{Table-dv}Difference Between the Euclidean Distance and $d_V(\cdot,\cdot)$.}
\centering
\begin{tabular}{c|c c c c c c} \hline\hline
Distribution & $V(\mu)$ & $\mu_1$ & $\mu_2$ & $Y$ &  $\frac{(Y-\mu_2)^2}{(Y-\mu_1)^2}$ & $\frac{d_V(Y,\mu_2)}{d_V(Y,\mu_1)}$  \\ \hline
Binomial & $\mu(1-\mu)$ & $.5$ & $.75$ & $1$ & .25 & .2991  \\  
Poisson & $\mu$ & 1 & 2 & 3 & .25 & .25 \\
Gamma & $\mu^2$ & 1 & 2 & 3 & .25 & .3805 \\
Inverse Gaussian & $\mu^3/2$ & 0 & 1 & 2 & .25 & .6735 \\ \hline\hline
\end{tabular}
\end{table}
\end{center}

Our definition of $R^2$ for generalized linear mixed models will proceed by replacing the Euclidean distance by the above manifold distance along the variance function.

Replacing $(a-b)^2$ with $d_V(a, b)$ in (\ref{eqn-lmmr2f}), we can define the proportion of variation in $Y_{ij}$ modeled by the fixed effects as
\begin{eqnarray} \label{eqn-glmmr2fr}
\rho^2_F = 1-\frac{E[d_V(Y_{ij}, E[Y_{ij}\mid \eta^F_{ij}])]}{E[d_V(Y_{ij}, E[Y_{ij}])]}.
\end{eqnarray}
Note that $E[Y_{ij}\mid \eta^F_{ij}]$ is estimated via $\hat{\eta}^F_{ij}$ resulted from fitting the linear mixed model in (\ref{eqn-lmmr2fest}). Suppose $\hat{\eta}^F_{ij}$ can also be obtained by fitting the generalized linear mixed model (\ref{eqn-glmm}). However, unlike the linear mixed models, fitting a generalized linear model with only the fixed effect $\eta^F_{ij}$ but ignoring the random effect $\eta^R_{ij}$ in model (\ref{eqn-glmm}) may result in an estimate $\tilde{\eta}^F_{ij}$ which is much different from $\hat{\eta}^F_{ij}$. Since $\tilde{\eta}^F_{ij}$, instead of $\hat{\eta}^F_{ij}$, represents better the contribution of fixed effects to explaining the variation in the response variable, we propose to estimate $\rho_F^2$ by
\begin{eqnarray} \label{eqn-glmmr2fest}
R_F^2 = 1-\frac{\sum_{i,j} d_V(Y_{ij},g^{-1}(\tilde{\eta}^F_{ij}))}{\sum_{i,j} d_V(Y_{ij}, \bar{Y}_{\cdot\cdot})}.
\end{eqnarray}

With (\ref{eqn-lmmr2fr}), we can similarly define the proportion of variation in $Y_{ij}$ modeled by both fixed and random effects as
\begin{eqnarray} \label{eqn-glmmr2mr}
\rho_M^2 = 1-\frac{E[d_V(Y_{ij}, E[Y_{ij}\mid \eta^F_{ij}, \eta^R_{ij}])]}{E[d_V(Y_{ij}, E[Y_{ij}])]} = 1-\frac{E[d_V(Y_{ij}, g^{-1}(\eta^F_{ij}+\eta^R_{ij}))]}{E[d_V(Y_{ij}, E[Y_{ij}])]}.
\end{eqnarray}
We can simply plug in $\hat{\eta}^F_{ij}$ and $\hat{\eta}^R_{ij}$, estimates of $\eta^F_{ij}$ and $\eta^R_{ij}$ respectively from fitting (\ref{eqn-glmm}), to estimate $\rho_M^2$,
\begin{eqnarray} \label{eqn-glmmr2mest}
R_M^2 = 1-\frac{\sum_{i,j} d_V(Y_{ij}, g^{-1}(\hat{\eta}^F_{ij}+\hat{\eta}^R_{ij}))}{\sum_{i,j} d_V(Y_{ij}, \bar{Y}_{\cdot\cdot})}.
\end{eqnarray}


The proportion of variation in $Y_{ij}$ modeled by random effects can be defined as
\begin{eqnarray} \label{eqn-glmmr2r}
\rho_{R}^2 = \rho_M^2 - \rho_{F}^2 = \frac{E[d_V(Y_{ij},E[Y_{ij}\mid \eta^F_{ij}])] - E[d_V(Y_{ij},E[Y_{ij}\mid \eta^F_{ij}, \eta^R_{ij}])]}{\mbox{var}(Y_{ij})}.
\end{eqnarray}
Accordingly, we can calculate the $R^2$ for random effects using $R_R^2 = R_M^2-R_F^2$ with $R_M^2$ and $R_F^2$ calculated in (\ref{eqn-glmmr2mest}) and (\ref{eqn-glmmr2fest}), respectively. Such a calculation of $R_R^2$ based on the above definition of $\rho_{R}^2$ follows the fact that we usually want to evaluate the contribution due to random effects by removing those attributable to the fixed effects. However, in the case that we prioritize the contribution due to the random effects over the contribution due to the fixed effects, we may parallel the definition of $\rho_{R}^2$ and $R_R^2$, instead of $\rho_{F}^2$ and $R_F^2$, to that of $\rho_{M}^2$ and $R_M^2$.

Similar to classical $R^2$ defined for linear regression models, both $R_F^2$ and $R_M^2$ defined as above may increase as more predictors are included into the underlying model. We can define the adjusted $R_F^2$ and adjusted $R_M^2$ by dividing the numerators and denominators in (\ref{eqn-glmmr2fest}) and (\ref{eqn-glmmr2mest}) with proper degrees of freedom. The adjusted $R_R^2$ can be accordingly calculated via the adjusted $R_F^2$ and adjusted $R_M^2$.

\section{Simulation Studies}


\subsection{Linear Mixed Models}

For linear mixed models, we compared the performance of our proposed $R^2$ to those proposed by \citet{Xu2003} and \citet{Nakagawa2017}, with a total of 1,000 data sets simulated from each model under investigation. Each data set has a total of 200 random samples, evenly clustered inside $m$ groups with $m=10$ and $50$, respectively. A binary covariate $X_1$ was generated for each observation, with half observations within the same group taking 1 and the other half taking -1. A second variable $X_2$ was generated from the standard normal distribution, independent of $X_1$ and the response variable. The $j$-th response value inside the $i$-th cluster, i.e., $y_{ij}$, was generated by
\[
y_{ij} = \mu_i + x_{1ij}\beta + \epsilon_{ij},
\]
where $x_{1ij}$ is the corresponding value of the binary covariate $X_1$, the random effect $\mu_i\stackrel{iid}{\sim} N(0,1)$, and $\epsilon_{ij}\stackrel{iid}{\sim} N(0,1)$.

\begin{figure}[htbp]
\begin{minipage}[h]{0.495\linewidth}
\centering a. $\rho_M^2$ when $m=10$
\end{minipage}
\begin{minipage}[h]{0.495\linewidth}
\centering b. $\rho_M^2$ when $m=50$
\end{minipage}
\begin{minipage}[h]{0.495\linewidth}
\centering
\includegraphics[width=3in, height=2.5in, clip=true]{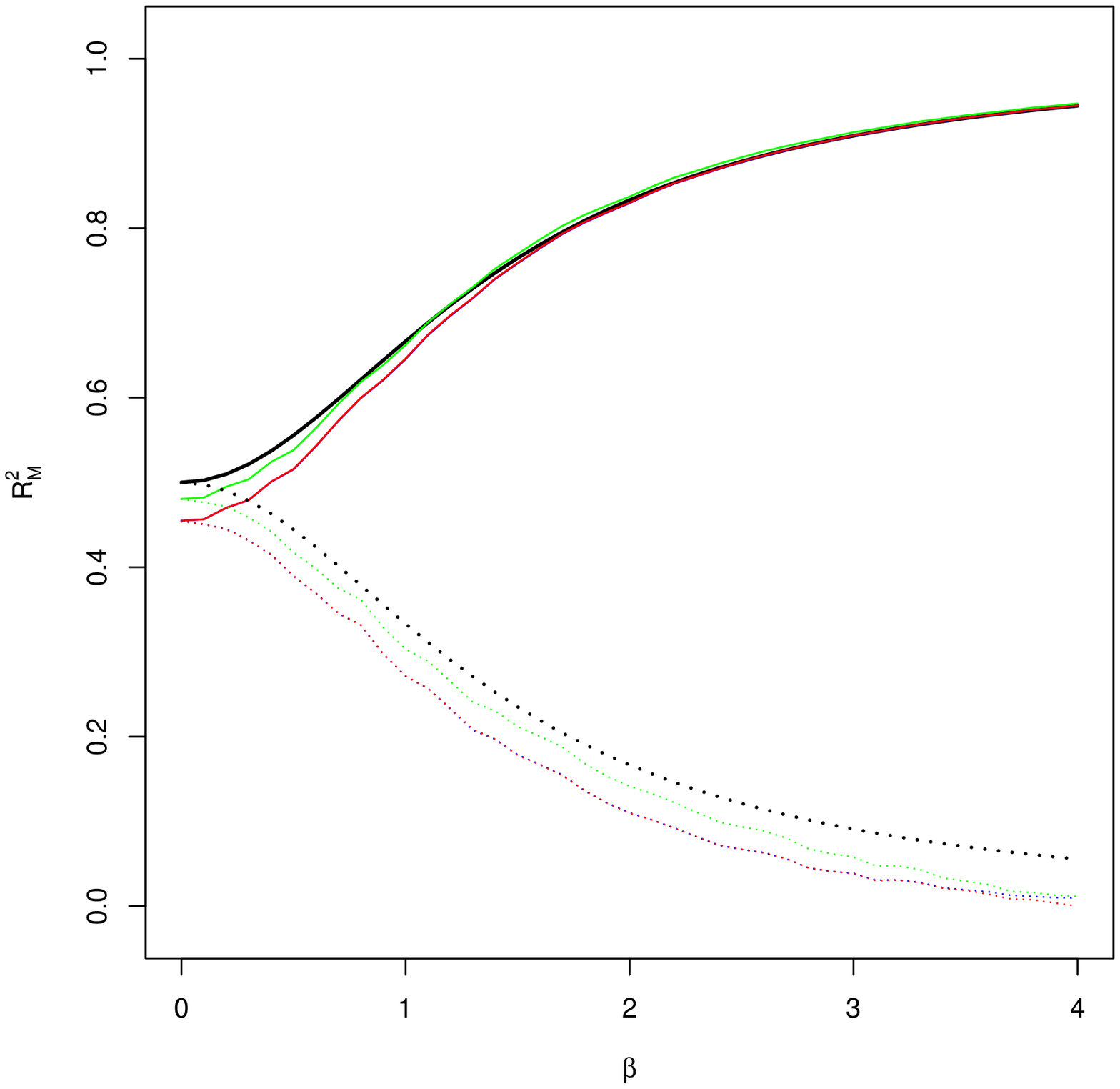}
\end{minipage}
\begin{minipage}[h]{0.495\linewidth}
\centering
\includegraphics[width=3in,height=2.5in,clip=true]{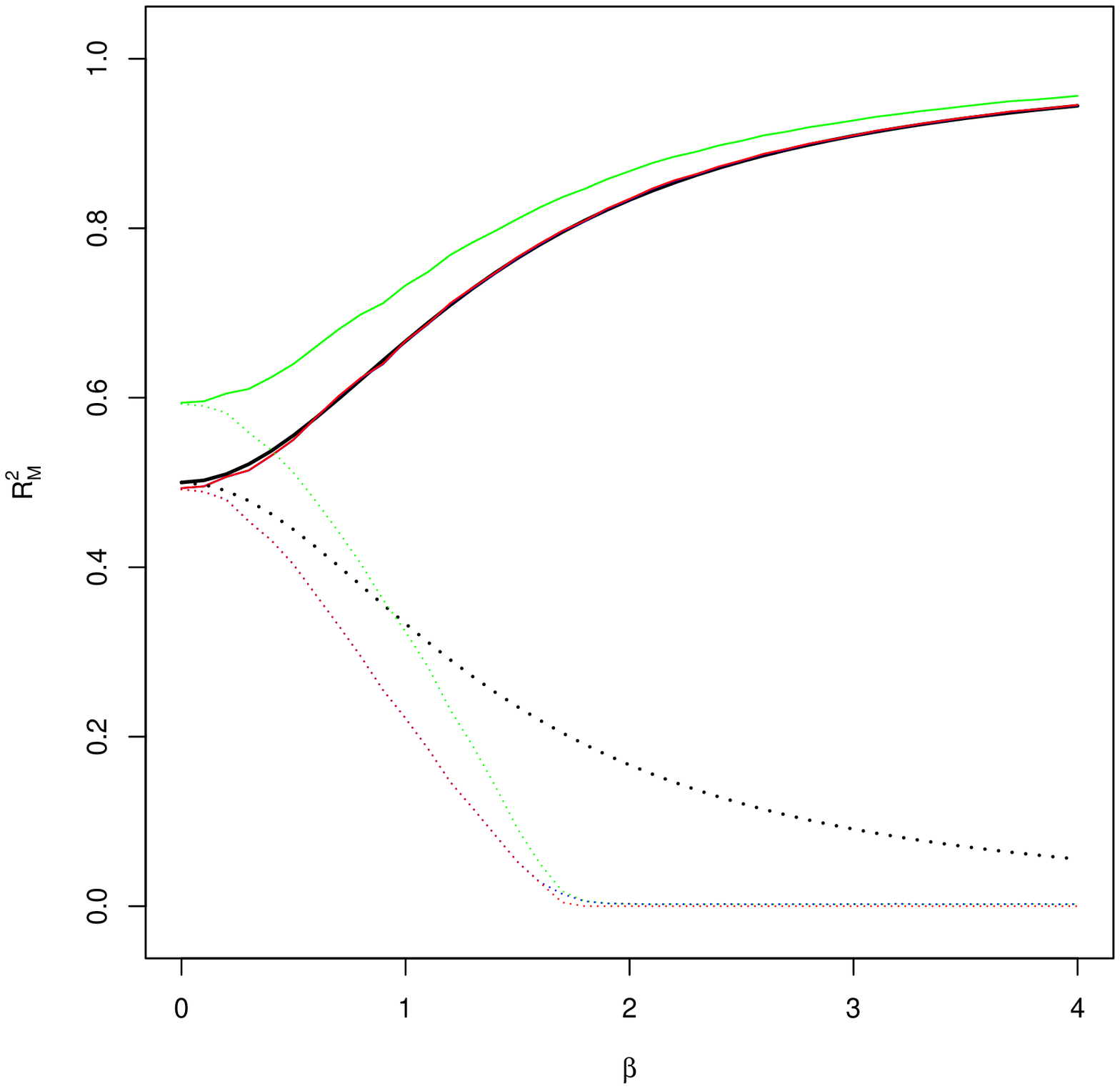}
\end{minipage}

\vskip6pt
\begin{minipage}[h]{0.495\linewidth}
\centering c. $\rho_F^2$ when $m=10$
\end{minipage}
\begin{minipage}[h]{0.495\linewidth}
\centering d. $\rho_F^2$ when $m=50$
\end{minipage}
\begin{minipage}[h]{0.495\linewidth}
\centering
\includegraphics[width=3in,height=2.5in,clip=true]{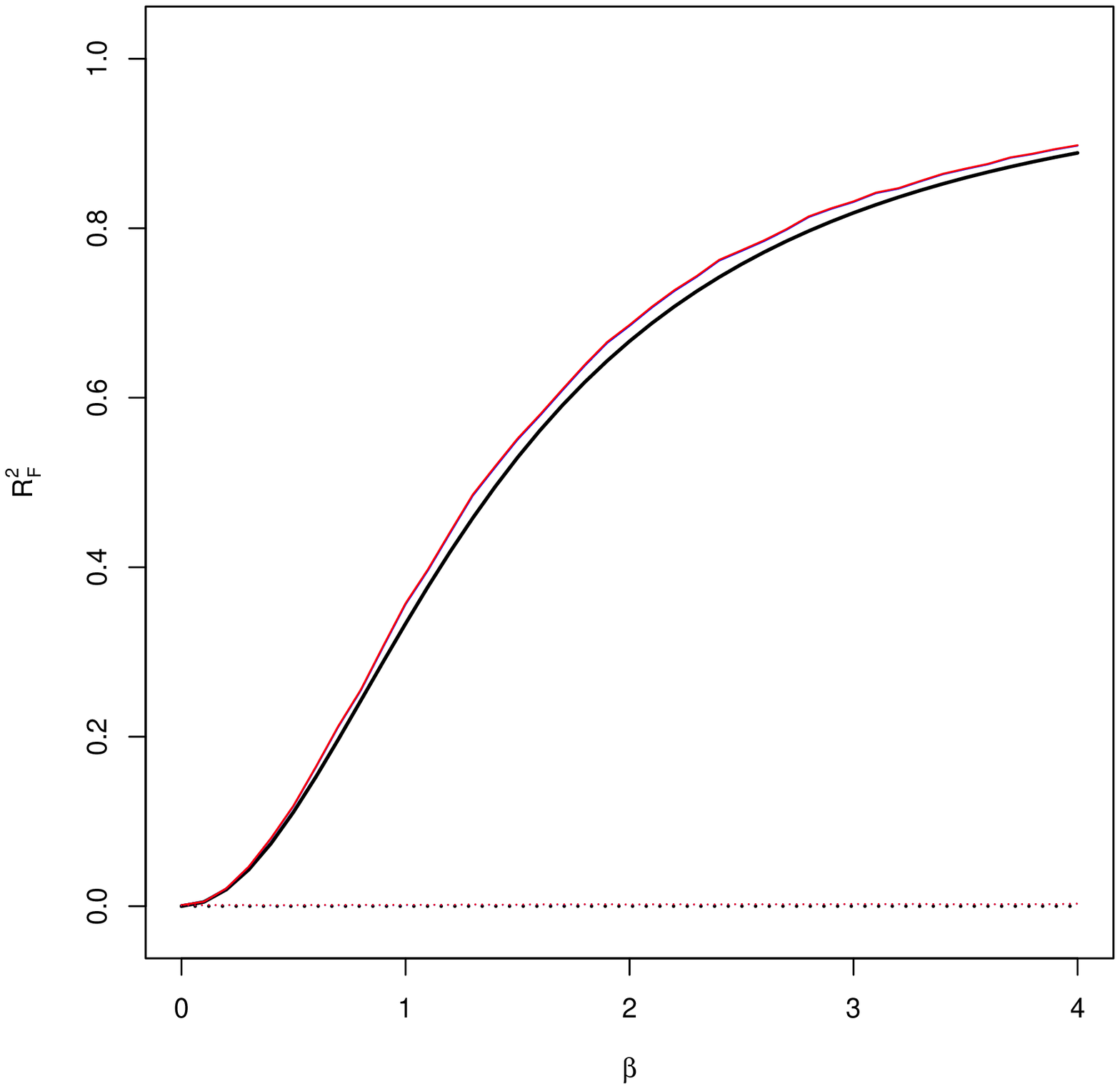}
\end{minipage}
\begin{minipage}[h]{0.495\linewidth}
\centering
\includegraphics[width=3in,height=2.5in,clip=true]{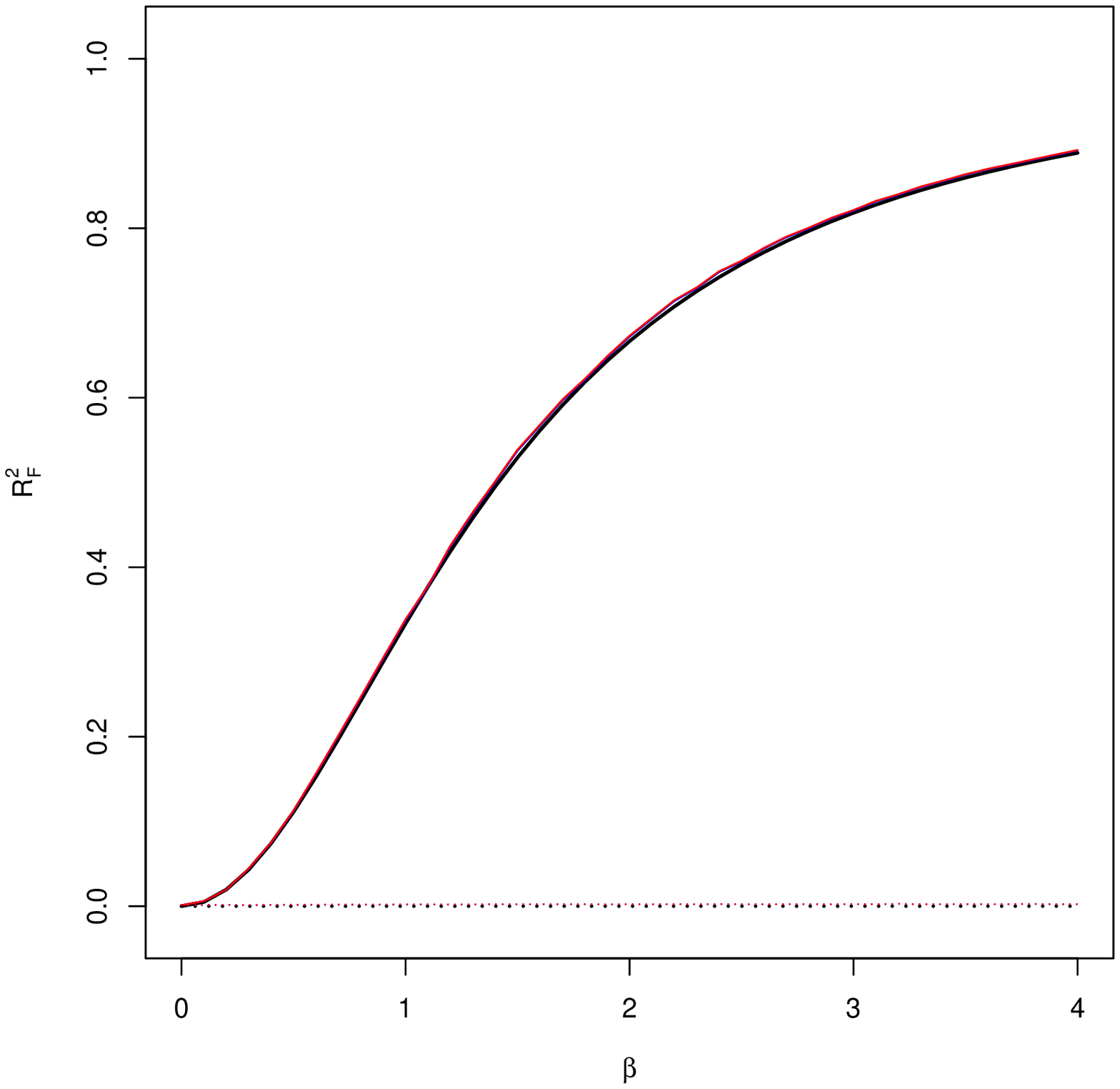}
\end{minipage}

\caption{\label{Figure-R2Gaussian}Medians of estimated $\rho_M^2$ and $\rho_F^2$ in linear mixed models (with $\rho_M^2$ and $\rho_F^2$ shown in black). Shown in the plot are $R_M^2$ and $R_F^2$ (blue), the method by \citet{Nakagawa2017} (red), and the method by \citet{Xu2003} (green). The solid lines are for the models including $X_1$, and the dashed lines are for the models including $X_2$ instead.}
\end{figure}

For each data set, we fit different models, i.e., regressing against only $X_1$ or only $X_2$, with the maximum likelihood method, and then estimate both $\rho_M^2$ and $\rho_F^2$ with different approaches as shown in Figure~\ref{Figure-R2Gaussian}. Note that, when regressing against only $X_1$, we have
\[
\rho^2_M = \frac{\beta^2+1}{\beta^2+2}, \ \ \ \ \ \ \rho^2_F = \frac{\beta^2}{\beta^2+2}.
\]
While \citet{Xu2003} only estimates $\rho_M^2$ but not $\rho_F^2$, it may severely overestimate $\rho_M^2$ especially when the number of groups is large. Indeed, the bias can be more than $0.15$ when there are 100 groups (the results are not shown). Our proposed $R_M^2$ and the methods by \citet{Nakagawa2017} estimate $\rho_M^2$ and $\rho_F^2$ well, however, there are slight bias especially when the number of groups is small. When regressing against only $X_2$, we have
\[
\rho^2_M = \frac{1}{\beta^2+2}, \ \ \ \ \ \ \rho^2_F = 0.
\]
While our proposed $R_F^2$ and the method by \citet{Nakagawa2017} estimate $\rho^2_F$ very well, they both underestimate $\rho^2_M$, which may be a desirable property for model selection.

We also fitted all models using the restricted maximum likelihood method and resultant coefficients of determinations are in general very close to the corresponding values calculated from the maximum likelihood estimators (the results are not shown). In particular, the method by \citet{Xu2003} is rarely affected by the choice of the restricted maximum likelihood or maximum likelihood methodestimator because the difference between restricted maximum likelihood and maximum likelihood estimators lies only in the estimated variances of the random effects, i.e., $\tau_{ij}^2$. 



\subsection{Logistic Mixed Models}


For logistic models, we compared the performance of our proposed $R^2$ to those proposed by \citet{Nakagawa2017}, with a total of 1,000 data sets simulated from each model under investigation. Each data set has a total of 400 random samples, evenly clustered inside $m$ groups with $m=10$ and $50$, respectively. The binary $X_1$ and continuous $X_2$ were generated similarly to those in the previous section. The $j$-th response value inside the $i$-th cluster, i.e., $y_{ij}$, was generated from a Bernoulli distribution with
\[
E[y_{ij}\mid \mu_i,x_{1ij}] = \left\{1+\exp(-\mu_i - x_{1ij}\beta)\right\}^{-1},
\]
where $x_{1ij}$ is the corresponding value of the binary covariate $X_1$, and the random effect $\mu_i\stackrel{iid}{\sim} N(0,1)$.

\begin{figure}[htbp]
\begin{minipage}[h]{0.495\linewidth}
\centering a. $\rho_M^2$ when $m=10$
\end{minipage}
\begin{minipage}[h]{0.495\linewidth}
\centering b. $\rho_M^2$ when $m=50$
\end{minipage}
\begin{minipage}[h]{0.495\linewidth}
\centering
\includegraphics[width=3in,height=2.5in,clip=true]{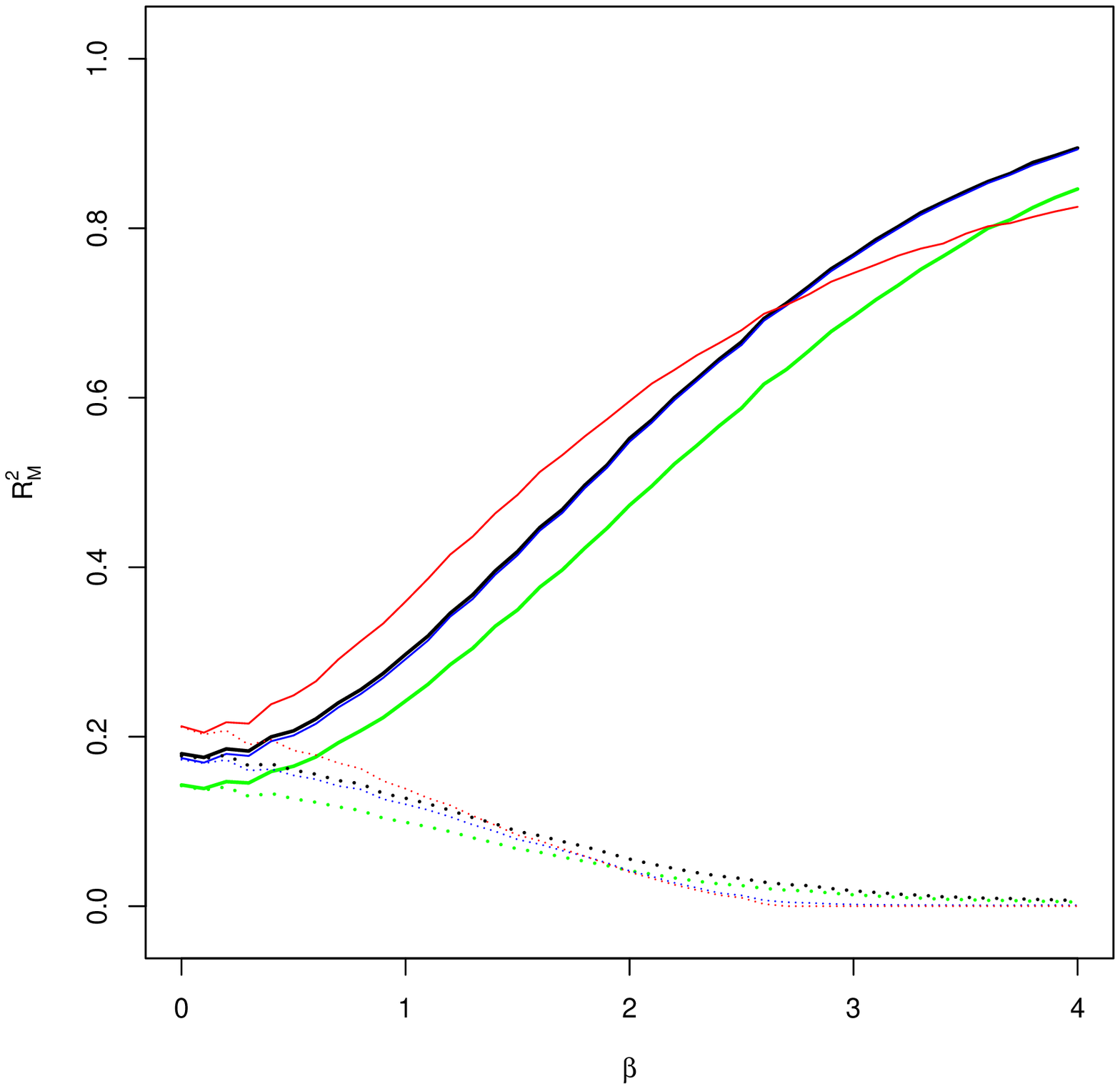}
\end{minipage}
\begin{minipage}[h]{0.495\linewidth}
\centering
\includegraphics[width=3in,height=2.5in,clip=true]{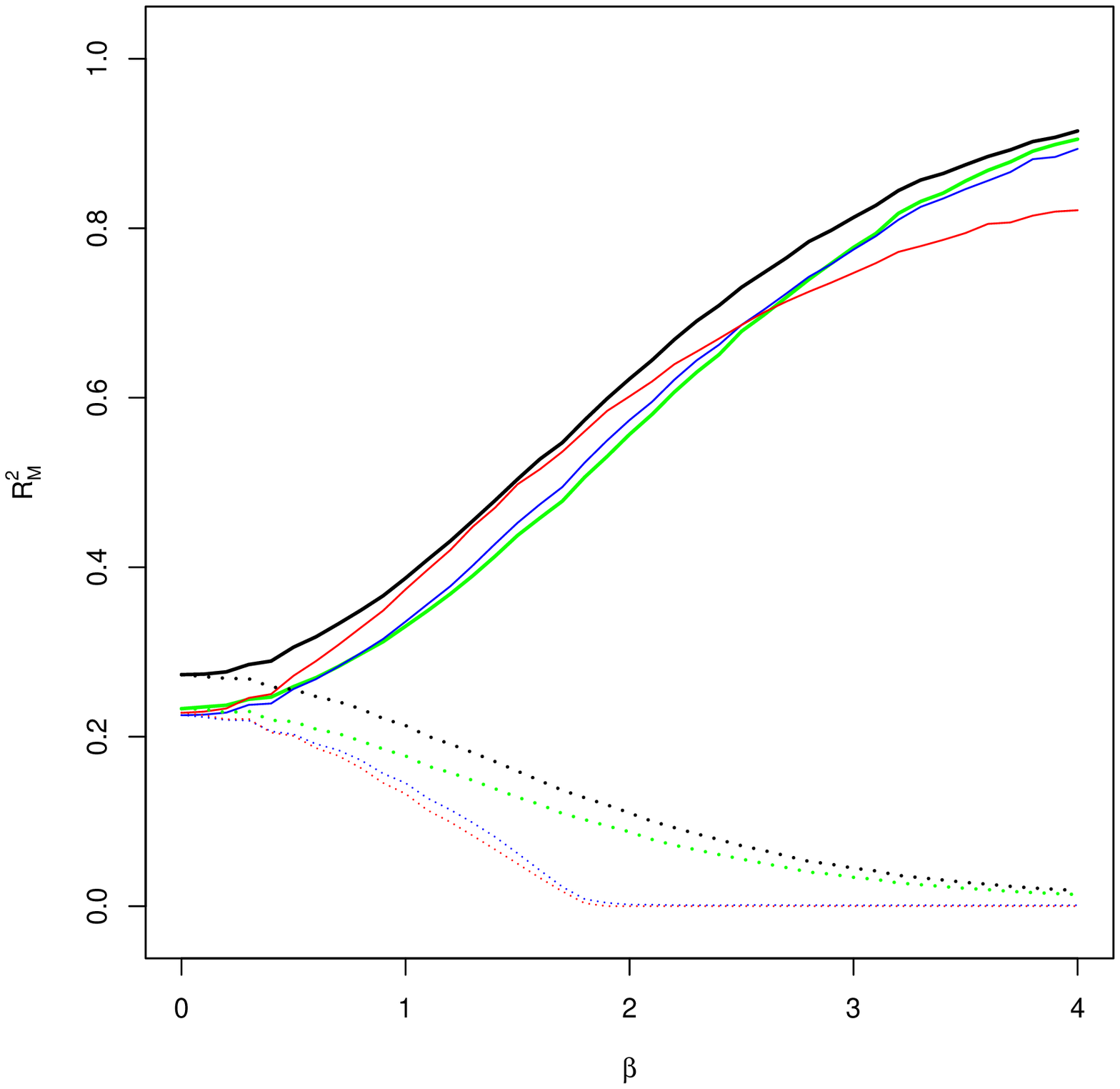}
\end{minipage}

\vskip6pt
\begin{minipage}[h]{0.495\linewidth}
\centering c. $\rho_F^2$ when $m=10$
\end{minipage}
\begin{minipage}[h]{0.495\linewidth}
\centering d. $\rho_F^2$ when $m=50$
\end{minipage}
\begin{minipage}[h]{0.495\linewidth}
\centering
\includegraphics[width=3in,height=2.5in,clip=true]{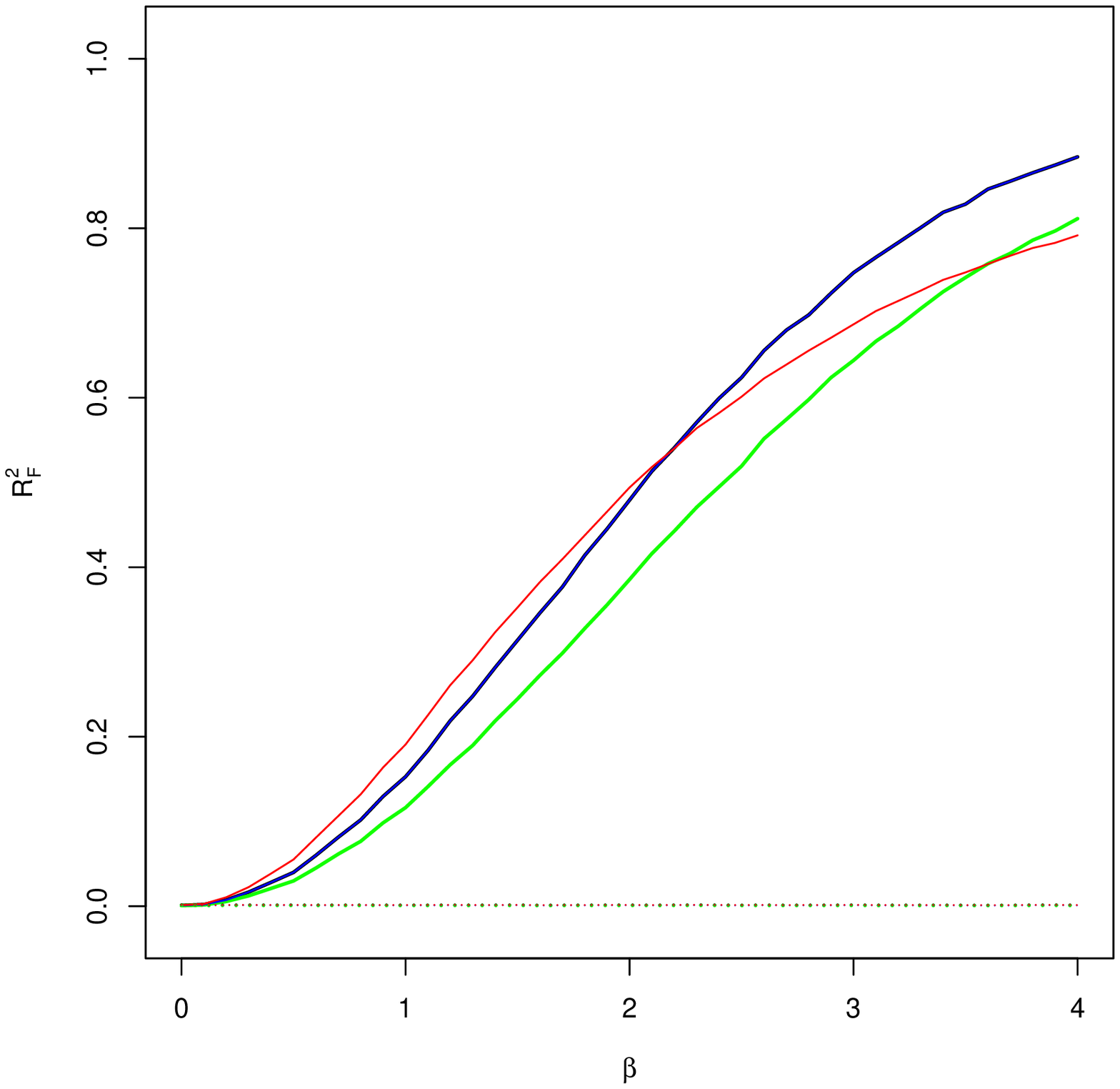}
\end{minipage}
\begin{minipage}[h]{0.495\linewidth}
\centering
\includegraphics[width=3in,height=2.5in,clip=true]{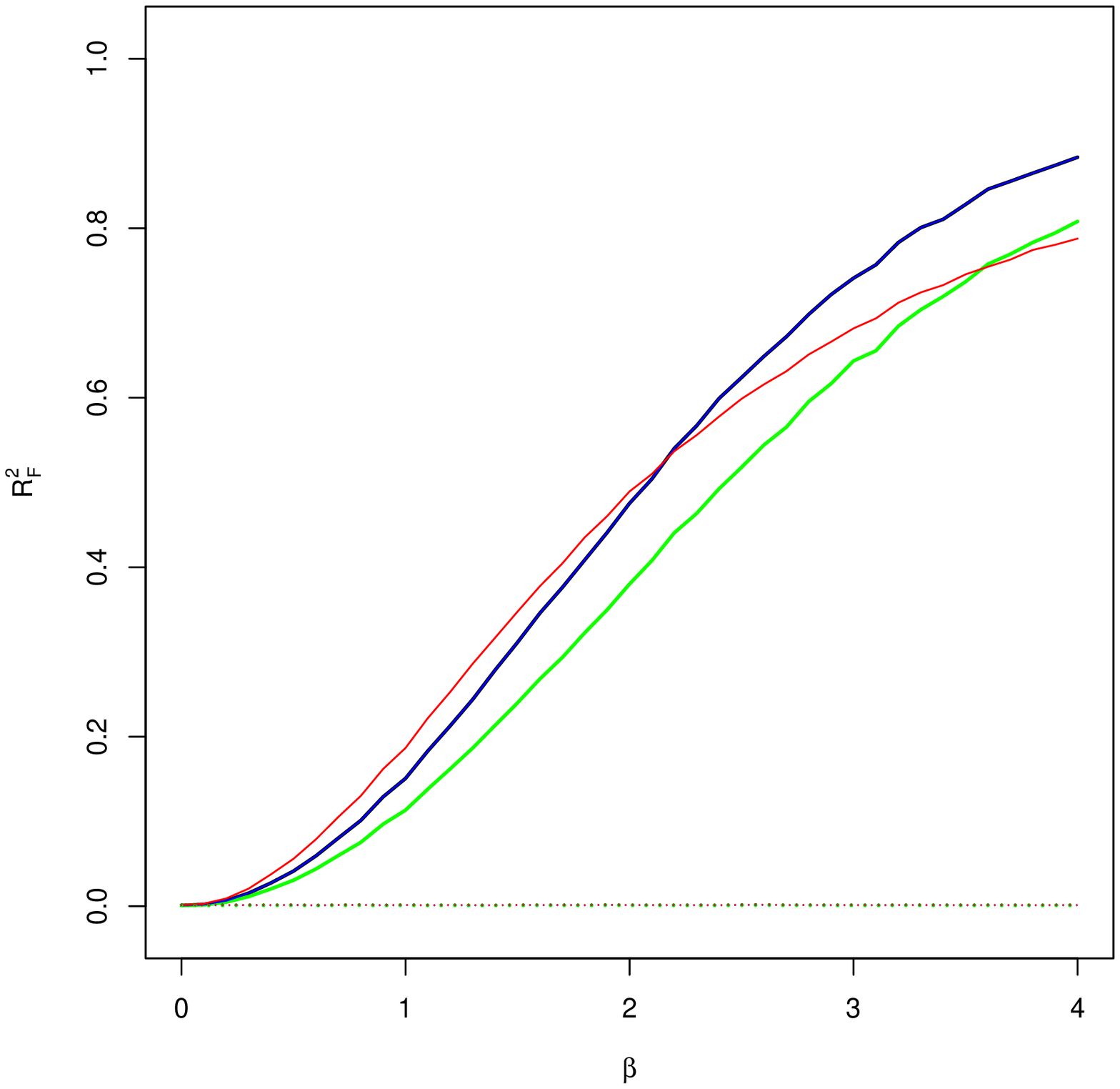}
\end{minipage}

\caption{\label{Figure-R2Binomial}Medians of estimated $\rho_M^2$ and $\rho_F^2$ in logistic mixed models. Shown in the plot are $R_M^2$ and $R_F^2$ (blue), and the method by \citet{Nakagawa2017} (red). For reference, we also plot $R_{KL}^2$ (green) and $R_V^2$ (black) for corresponding generalized linear models. The solid lines are for the models including $X_1$, and the dotted lines are for the models including $X_2$ instead.}
\end{figure}

For each data set, we fit different models with the maximum likelihood method, and then estimate both $\rho_M^2$ and $\rho_F^2$ with our approach and the method by \citet{Nakagawa2017} as shown in Figure~\ref{Figure-R2Binomial}. With true values unknown, we fit a generalized linear model for each mixed-effects model by including fixed, instead of random, group effects, and calculated $R_{KL}^2$ by \citet{Cameron1997} and $R_V^2$ by \citet{Zhang2017}. Note that $R_V^2$ may conceptually favor our proposed $R^2$. However, $R_{KL}^2$ can serve well as a benchmark although $R_{KL}^2$ takes into account of whole distributional diffusion including quadratic variation. Figure~\ref{Figure-R2Binomial} indeed shows that both $R_V^2$ and $R_{KL}^2$ have similar increasing patterns with the discrepancies due to their conceptual difference.

When the true predictor $X_1$ is included in the model, the estimated $\rho_M^2$ and $\rho_F^2$ by \citet{Nakagawa2017} are in general flatter than other estimates. All estimates of $\rho_M^2$ and $\rho_F^2$ are slightly affected by the number of groups. It is not surprising that, in Figures~\ref{Figure-R2Binomial}.a and \ref{Figure-R2Binomial}.b, both $R_{KL}^2$ and $R_V^2$ are significantly affected by the number of groups as including more fixed group effects in the model may inflate the corresponding $R^2$. Both $R_{KL}^2$ and $R_V^2$ perform consistently between Figures~\ref{Figure-R2Binomial}.c and \ref{Figure-R2Binomial}.d as the group effects are excluded from the models. Note that $R_F^2$ and $R_V^2$ are overlapped in Figures~\ref{Figure-R2Binomial}.c and \ref{Figure-R2Binomial}.d due to their identity by definition. When regressing $Y$ vs. $X_2$ by excluding the true predictor $X_1$, our proposed measures and the measures by \citet{Nakagawa2017} perform consistently with $m=10$ and $m=50$. In general, $R_M^2$ and $R_F^2$ capture well the increasing proportion of explained variation in a binary response variable, and perform better than the method by \citet{Nakagawa2017}.



\subsection{Loglinear Mixed Models}

For loglinear models, we compared the performance of our proposed $R^2$ to those proposed by \citet{Nakagawa2017}, with a total of 1,000 data sets simulated from each model under investigation. Each data set has a total of 400 random samples, evenly clustered inside $m=50$ groups. The binary $X_1$ and continuous $X_2$ were generated similarly to those in the previous sections. The $j$-th response value inside the $i$-th cluster, i.e., $y_{ij}$, was generated from a Poisson distribution with
\[
E[y_{ij}\mid \mu_i,x_{1ij}] = \exp(\mu_i + x_{1ij}\beta),
\]
where $x_{1ij}$ is the corresponding value of the binary covariate $X_1$, and the random effect $\mu_i\stackrel{iid}{\sim} N(0,.25)$.
We also simulated overdispersed counts from a negative binomial distribution with the success probability
\[
p_{ij}  = 1/\left\{1+\exp(-\mu_i - x_{1ij}\beta)\right\}.
\]

\begin{figure}[htbp]
\begin{minipage}[h]{0.495\linewidth}
\centering a. $\rho_M^2$ for Poisson Counts
\end{minipage}
\begin{minipage}[h]{0.495\linewidth}
\centering b. $\rho_M^2$ for Overdispersed Counts
\end{minipage}
\begin{minipage}[h]{0.495\linewidth}
\centering
\includegraphics[width=3in,height=2.5in,clip=true]{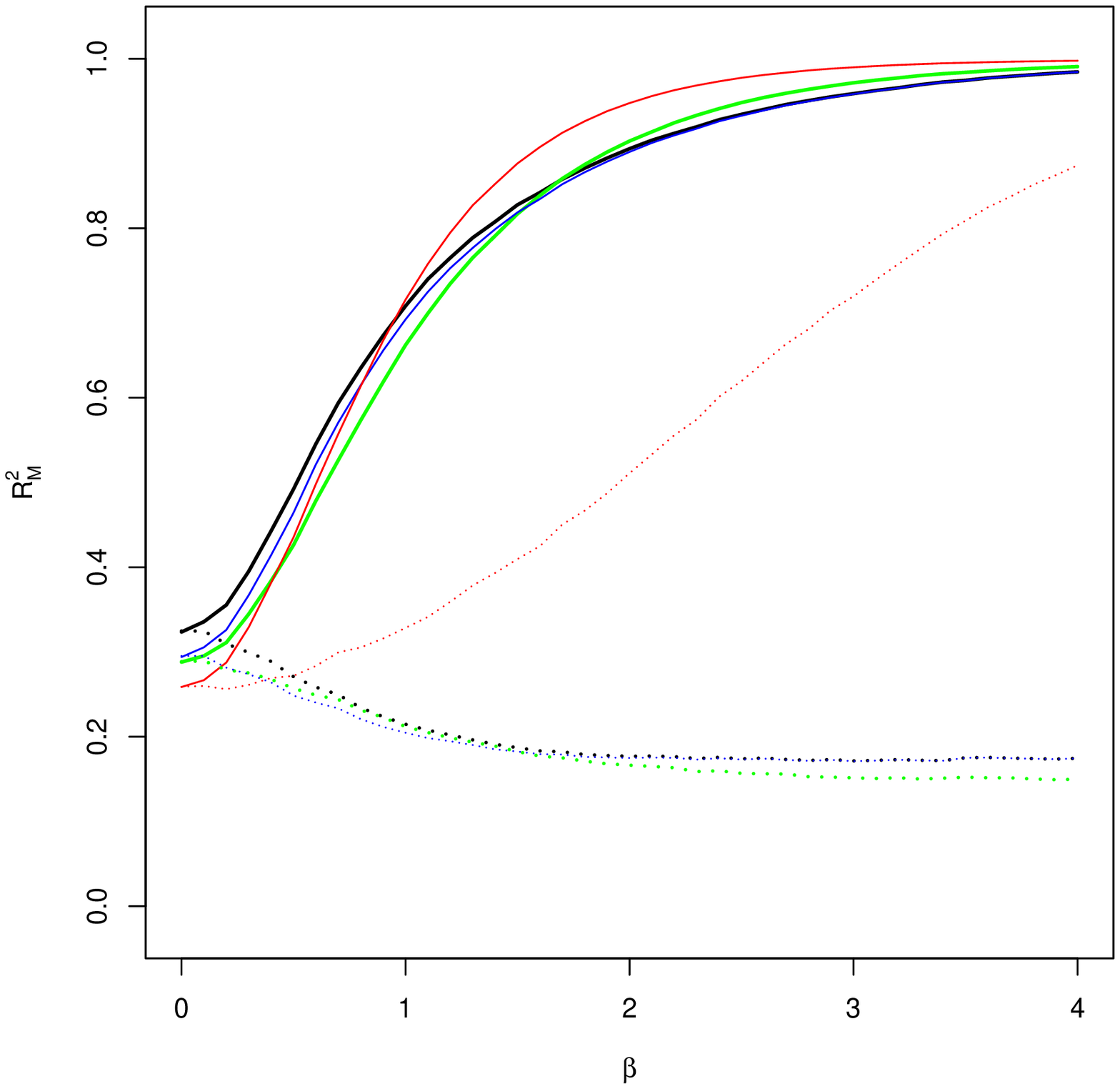}
\end{minipage}
\begin{minipage}[h]{0.495\linewidth}
\centering
\includegraphics[width=3in,height=2.5in,clip=true]{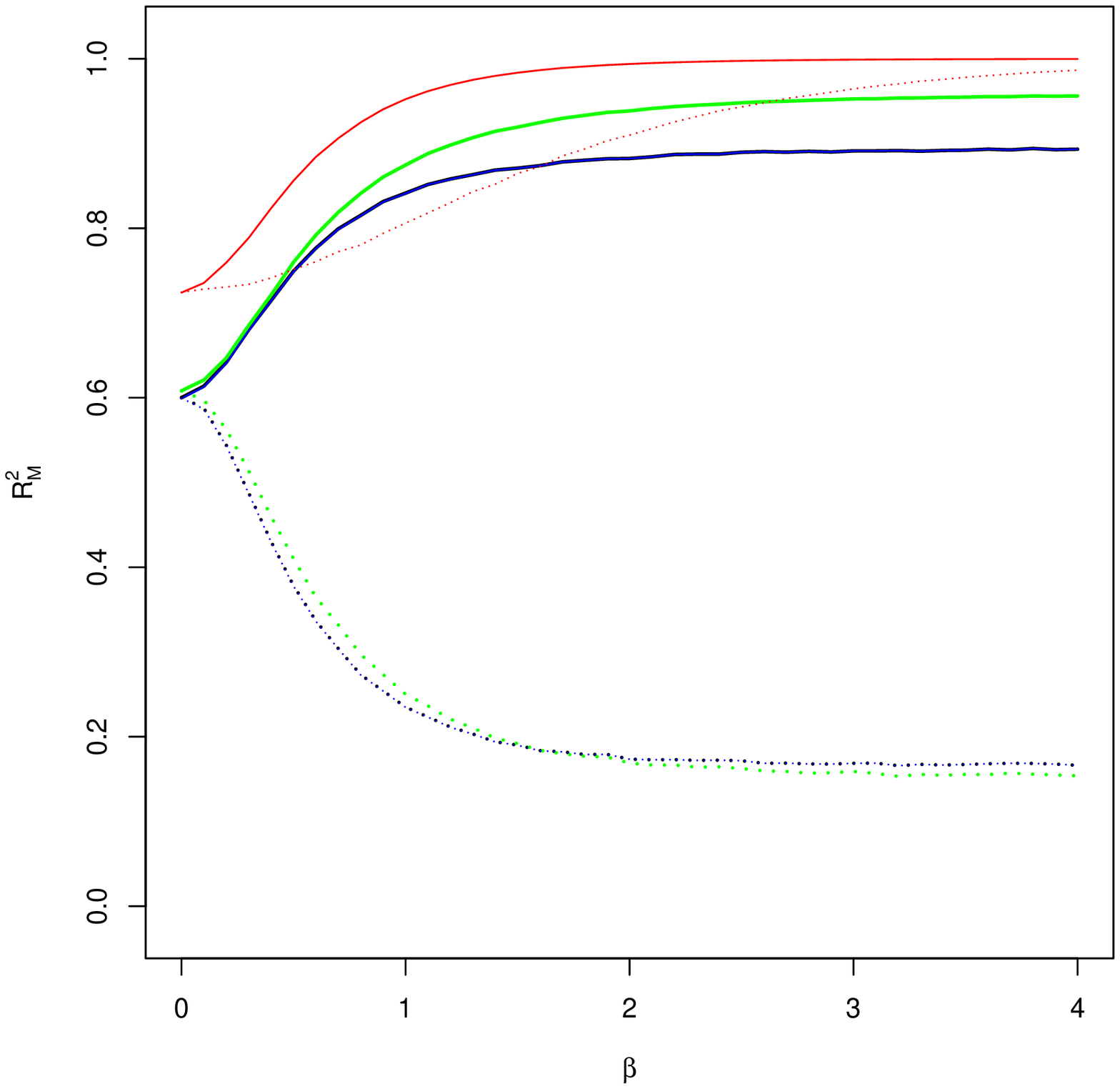}
\end{minipage}

\vskip6pt
\begin{minipage}[h]{0.495\linewidth}
\centering c. $\rho_F^2$ for Poisson Counts
\end{minipage}
\begin{minipage}[h]{0.495\linewidth}
\centering d. $\rho_F^2$ for Overdispersed Counts
\end{minipage}
\begin{minipage}[h]{0.495\linewidth}
\centering
\includegraphics[width=3in,height=2.5in,clip=true]{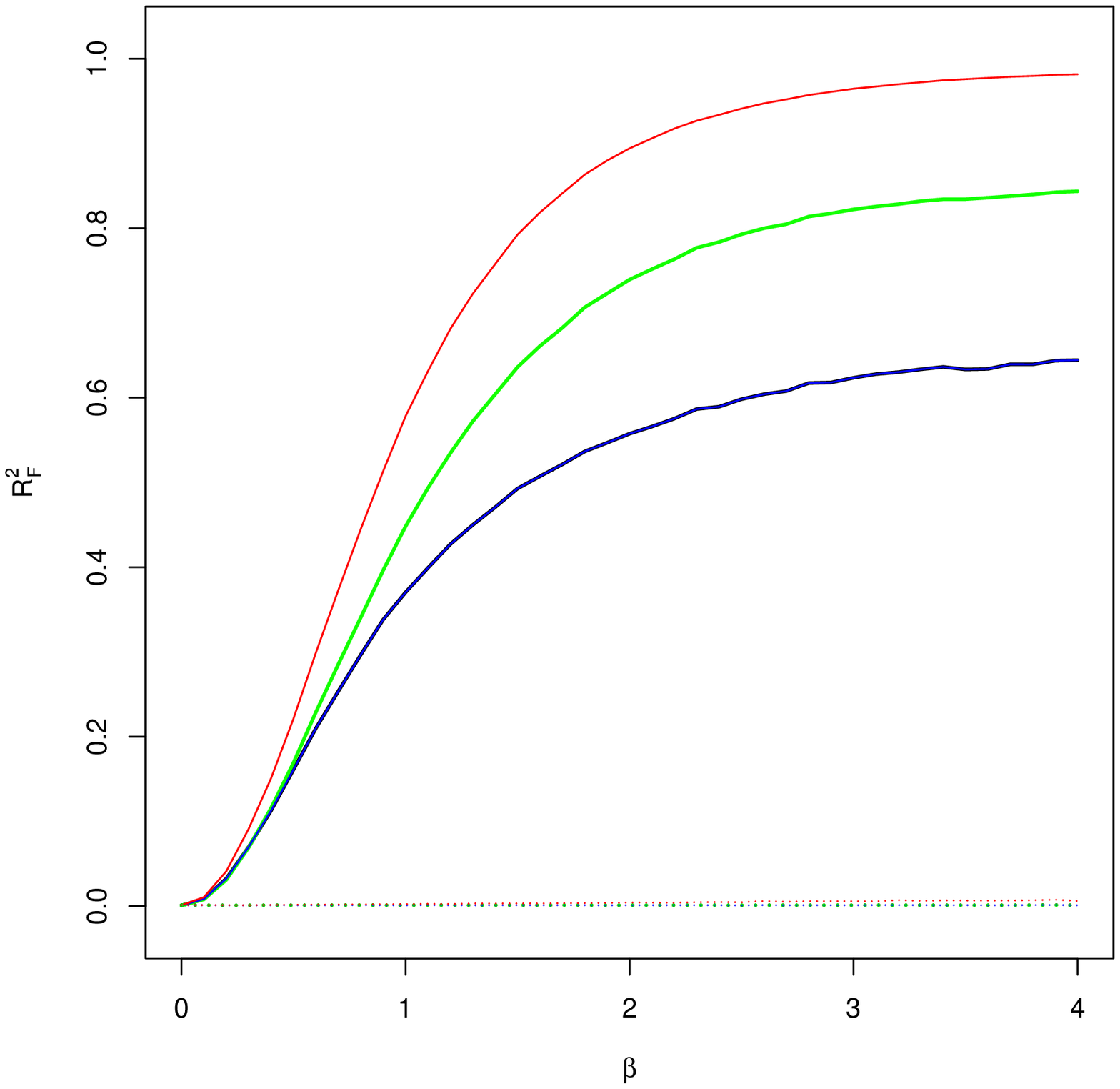}
\end{minipage}
\begin{minipage}[h]{0.495\linewidth}
\centering
\includegraphics[width=3in,height=2.5in,clip=true]{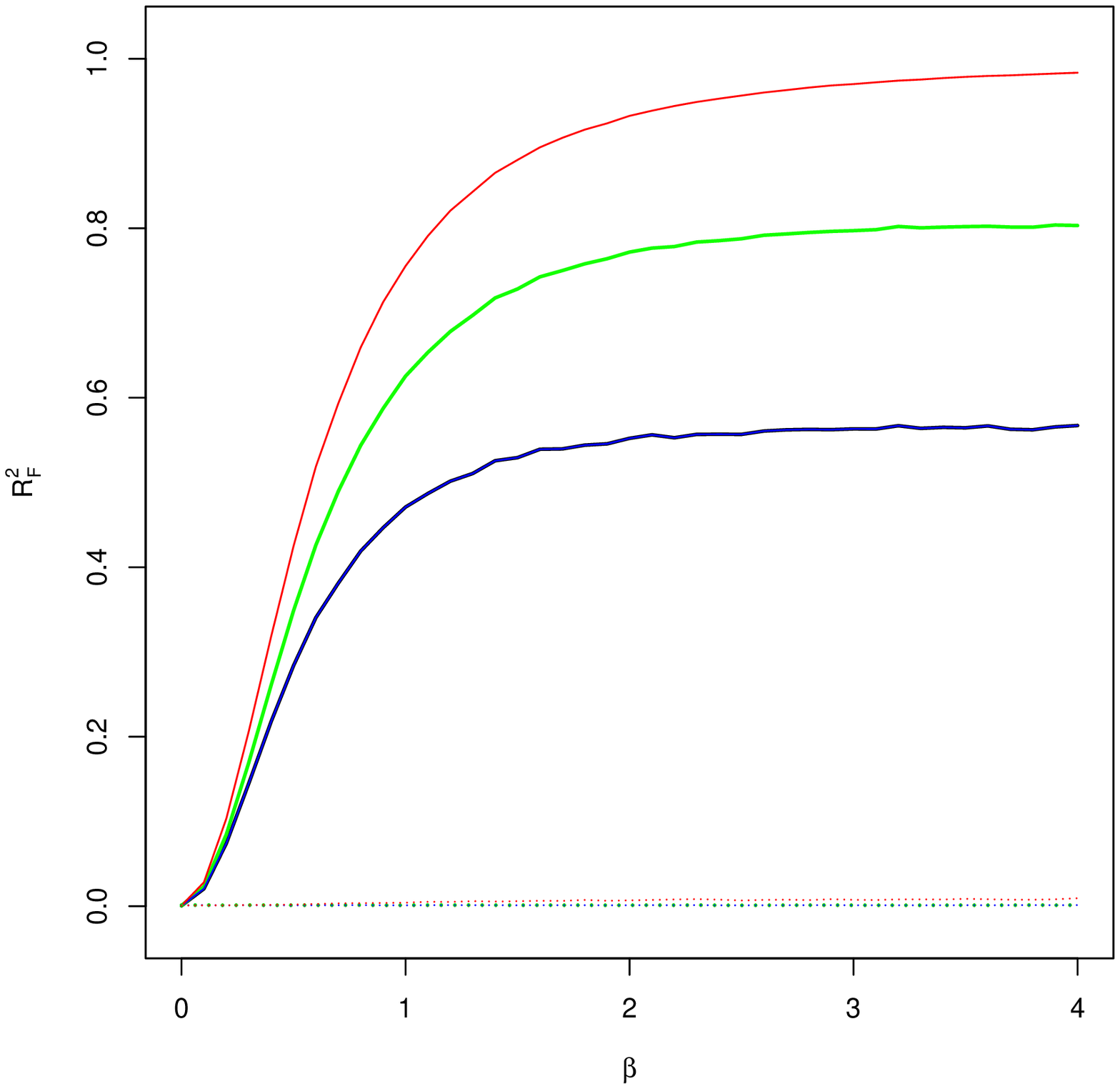}
\end{minipage}

\caption{\label{Figure-R2Poisson}Medians of estimated $\rho_M^2$ and $\rho_F^2$ in loglinear mixed models. Shown in the plot are $R_M^2$ and $R_F^2$ (blue), and the method by \citet{Nakagawa2017} (red). For reference, we also plot $R_{KL}^2$ (green) and $R_V^2$ (black) for corresponding generalized linear models. The solid lines are for the models including $X_1$, and the dotted lines are for the models including $X_2$ instead.}
\end{figure}

For each data set, we fit the different models with the maximum likelihood method, and then estimate both $\rho_M^2$ and $\rho_F^2$ with our approach and the method by \citet{Nakagawa2017} as shown in Figure~\ref{Figure-R2Poisson}. For each model, we also fit a generalized linear model by including fixed, instead of random, group effects, and calculated $R_{KL}^2$ by \citet{Cameron1997} and $R_V^2$ by \citet{Zhang2017} as benchmarks.

When the true predictor $X_1$ is included in the model, as shown in Figures~\ref{Figure-R2Poisson}.a and \ref{Figure-R2Poisson}.b, $R_M^2$ overlays its reference $R_V^2$ for the Poisson counts data but is slightly smaller than $R_V^2$ for the overdispersed counts data. As conceptually identical, it is not surprising to observe overlayed $R_V^2$ and $R_V^2$ in Figures~\ref{Figure-R2Poisson}.c and \ref{Figure-R2Poisson}.d. While $R_{KL}^2$ measures overall distributional diffusion including quadratic variation, it reports slightly larger values than $R_M^2$ in Figures~\ref{Figure-R2Poisson}.a and \ref{Figure-R2Poisson}.b but much larger values than $R_F^2$ in Figures~\ref{Figure-R2Poisson}.c and \ref{Figure-R2Poisson}.d, in particular when $\beta$ is large. On the other hand, the method by \citet{Nakagawa2017} reports slightly larger estimate of $\rho_M^2$ than $R_{KL}^2$ in Figures~\ref{Figure-R2Poisson}.a and \ref{Figure-R2Poisson}.b, but
much larger estimate of $\rho_F^2$ than $R_{KL}^2$ in Figures~\ref{Figure-R2Poisson}.c and \ref{Figure-R2Poisson}.d.

When $X_2$, instead of the true predictor $X_1$, is included in the model, $\rho_M^2$ only measures the proportion of variation due to the random effects so it should decrease when $\beta$ increases. While $R_M^2$ as well as $R_V^2$ and $R_{KL}^2$ decreases, the estimate by \citet{Nakagawa2017} increases and approach to one as $\beta$ increases, as shown in Figures~\ref{Figure-R2Poisson}.a and \ref{Figure-R2Poisson}.b. Therefore, the method by \citet{Nakagawa2017} falsely evaluates the proportion of variation explained by $X_2$. On the other hand, all methods perform well in estimating $\rho_F^2$ in this scenario as shown in Figures~\ref{Figure-R2Poisson}.c and \ref{Figure-R2Poisson}.d. In general, the method proposed by \citet{Nakagawa2017} tends to overstate the proportion of explained variation in a count response while $R_M^2$ and $R_F^2$ capture well the increasing pattern as benchmarked by $R_{KL}^2$.

\section{Real Data Analysis} \label{Sec-RealData}

\subsection{Corn Yield with Variable Nitrogen}

A total of 1,705 observations were collected in 2001 from four different topographic regions on corn yield with five different nitrogen treatments in Argentina \citep{Edmondson2014, Lambert2004}. Recorded with each observation is a brightness value, proxy for low organic matter content, which varies across different topographic regions. With possible fixed-effects predictors nitrogen treatments (N) and brightness value (B), we consider different linear mixed models, all having the topographic region as a factor with random effects. Since there is high association between brightness value and topographic region, we also regressed brightness value against topographic region and include its residual (B$_r$), instead of itself, in the models, see Table~\ref{Table-Corn}. As references, classical and adjusted coefficients of determination ($R^2$) are also calculated for the corresponding models but including the topographic region with fixed effects.

\begin{center}
\begin{table}[htbp]
\caption{\label{Table-Corn}Coefficients of Determination in Analysis of the Corn Yield Data$^{\ast}$}
\centering
\begin{tabular}{c | c c c c c c c c} \hline\hline
Fixed & \multicolumn{3}{c}{\underline{\ \ \ \ \ \ \ \ \ \ \ \ \ \ $\rho_M^2$\ \ \ \ \ \ \ \ \ \ \ \ \ \ }} & \multicolumn{2}{c}{\underline{\ \ \ \ \ $\rho_F^2$\ \ \ \ \ }} & & &  \\
Effects & $R_M^2$ & NJS$^a$ & Xu$^b$ & $R_F^2$ & NJS$^a$ & AIC & BIC & $R^2$ \\  \hline %
N, B, B$^2$ & .779({\bf .778}) & .754 & {\bf .768} & .303 & .089 & {\bf 13444} & 13499 & .769 ({\bf .767}) \\
N, B & .778 (.778) & .758 & .768 & .284 & .082 & 13446 & {\bf 13494} & .768 (.767) \\
N & .730 (.729) & {\bf .774} & .719 & .005 & .005 & 13770 & 13814 & .719 (.717) \\
B, B$^2$ & .774 (.774) & .748 & .763 & .297 & .083 & 13460 & 13498 & .763 (.762) \\
B & .774 (.773) & .753 & .762 & .279 & .076 & 13485 & 13507 & .762 (.762) \\ \hline
N, B$_r$, B$^2_r$ & .778 ({\bf .777}) & {\bf .813} & {\bf .770} & .069 & .047 & {\bf 13433} & {\bf 13488} & .770 ({\bf .769})  \\
N, B$_r$ & .776 (.775) & {\bf .813} & .768 & .054 & .044 & 13447 & 13496 & .768 (.767)  \\
B$_r$, B$^2_r$ & .773 (.773) & .809 & .764 & .063 & .042 & 13473 & 13501 & .764 (.764) \\
B$_r$ & .771 (.771) & .809 & .762 & .048 & .040 & 13486 & 13508 & .762 (.762) \\ \hline\hline %
\multicolumn{9}{l}{$^{\ast}$\footnotesize{With adjusted values in parentheses;} $^a$ \footnotesize{The method by \citet{Nakagawa2017};}} \\
\multicolumn{9}{l}{$^b$ \footnotesize{The method by \citet{Xu2003}.}}
\end{tabular}
\end{table}
\end{center}

When including original brightness value in the model, the method by \citet{Nakagawa2017} estimated $\rho_M^2$ with slightly lower values than $R_M^2$ except that, for the model with only N having fixed effects, the method by \citet{Nakagawa2017} reported its largest value at .774. Note that the method by \citet{Nakagawa2017} reported larger value for the model with B only then the model with B and B$^2$, and also decreasing values for the three increasing models: (i) the model with N only; (ii) the model with N and B; and (iii) the model with N, B, and B$^2$. Therefore, the method by \citet{Nakagawa2017} may not necessarily report larger values for larger models, unlike other unadjusted coefficients of determinations which usually report larger values for larger models. The method by \citet{Nakagawa2017} also estimated $\rho_F^2$ with much smaller values than $R_F^2$ except for the model with N only.


When the brightness values were replaced by by their residuals, the method by \citet{Nakagawa2017} estimated $\rho_M^2$ with larger values than $R_M^2$, and reported slightly smaller estimates of $\rho_F^2$ than $R_F^2$. While the method by \citet{Nakagawa2017} reported the largest estimate of $\rho_M^2$ at .813 for two sets of models including N and B$_r$. Note that the model with N, B$_r$, and B$_r^2$ is selected by AIC, BIC, adjusted $R^2$, the methods by \citet{Xu2003} and \citet{Nakagawa2017}, while adjusted $R_M^2$ reported this model with the highest value among all models replacing the brightness value with its residual. However, when selecting models by considering the original brightness values, the method by \citet{Nakagawa2017} selected the model with N only, while BIC selected the model with N and B, and all other methods selected the model with N, B, and B$^2$.


\subsection{Distribution of {\it Elaphostrongylus cervi} in Red Deers}


\citet{Vicente2006} studied the distribution of the first-stage larvae of {\it Elaphostrongylus cervi} in the red deer across Spain. A total of 826 deers across 24 farms were surveyed for the incidence \citep{Zuur2009}.  We considered different logistic mixed models to investigate how the length of a deer affected its chance to get the first-stage larvae of {\it Elaphostrongylus cervi}, and therefore include both the standardized length (L) and its quadratic term (L$^2$, as well as deer sex (S), as possible predictors. All mixed-effect models included farm as a factor with random effects, see Table~\ref{Table-Ecology}. As references, $R_V^2$ proposed by \citet{Zhang2017} and $R_{KL}^2$ proposed by \citet{Cameron1997} were calculated for corresponding models but including farm as a factor with fixed effects instead.

\begin{center}
\begin{table}[htbp]
\caption{\label{Table-Ecology}Coefficients of Determination in Analysis of the Red Deers Data$^{\ast}$}
\centering
\begin{tabular}{c|c c c c c c c c} \hline\hline
Fixed & \multicolumn{2}{c}{\underline{\ \ \ \ \ \ \ \ \ \ $\rho_M^2$\ \ \ \ \ \ \ \ \ \ }} & \multicolumn{2}{c}{\underline{\ \ \ \ \ $\rho_F^2$\ \ \ \ \ }} & & & & \\
Effects & $R_M^2$ & NJS$^a$ &  $R_F^2$ & NJS$^a$ & AIC & BIC & $R_V^2$ & $R_{KL}^2$\\  \hline %
S$\ast$L, S$\ast$L$^2$ & .352 ({\bf .347}) & .507 & .103 & .156 & 832 & 865 & .357 ({\bf .334}) & .308 ({\bf .284}) \\
S$\ast$L & .348 (.345) & .504 & .087 & .143 & 833 & 856 & .354 (.333) & .305 (.283) \\
S, L, L$^2$ & .349 (.346) & {\bf .509} & .102 & .099 & 831 & 854 & .354 (.333) & .306 (.283) \\
S, L & .340 (.337) & .501 & .083 & .131 & 841 & 860 & .345 (.324) & .296 (.274) \\
L, L$^2$ & .347 (.345) & .507 & .102 & .159 & {\bf 830} & {\bf 849} & .353 (.332) & .304 (.282) \\
L & .334 (.333) & .495 & .083 & .128 & 844 & 858 & .340 (.320) & .291 (.270) \\
S & .239 (.237) & .344 & .012 & .005 & 931 & 945 & .247 (.225) & .207 (.184) \\ \hline\hline %
\multicolumn{9}{l}{$^{\ast}$\footnotesize{With adjusted values in parentheses;} $^a$ \footnotesize{The method by \citet{Nakagawa2017}.}}
\end{tabular}
\end{table}
\end{center}

In general, the method by \citet{Nakagawa2017} provided slightly larger estimates of $\rho_F^2$ than $R_F^2$ (except two models with similar values), but much larger estimates of $\rho_M^2$ than $R_M^2$ in all models. On the other hand, $R_M^2$ reported very similar values to $R_V^2$ while $R_{KL}^2$ reported slightly smaller values than both of them. However, adjusted versions of $R_M^2$, $R_V^2$, and $R_{KL}^2$ all selected the largest model as the best, while both AIC and BIC selected the model with L and L$^2$, and the method by \citet{Nakagawa2017} selected the model with S, L, and L$^2$ instead.

While all of $R_M^2$, $R_V^2$, and $R_{KL}^2$ reported larger values for larger models, the method by \citet{Nakagawa2017} also reported larger values for larger models except the model with S, L, and L$^2$ for which it reported the largest value at .509, higher than the value for the largest model. However, both $R_F^2$ and the method by \citet{Nakagawa2017} estimated $\rho_F^2$ with larger values for larger models.

\subsection{The Begging Behavior of Nestling Barn Owls}

\citet{Roulin2007} studied vocal begging behaviour of nestling barn owls when the parents brought prey across 27 nests, with 2 to 7 nestlings per nest. A total of 599 observations collected with the response variable sibling negotiation, which is defined as the number of calls just before arrival of a parent at a nest, as well as the number of siblings per nest, the parent's sex (S), arrival time at the nest (T), and whether the nestings were food satiated or deprived (F) \citep{Zuur2009}. We considered loglinear mixed models with the number of calls offset by the number of siblings per nest. All models included nest as a factor with random effects, see Table~\ref{Table-Owl}. As references, $R_V^2$ proposed by \citet{Zhang2017} and $R_{KL}^2$ proposed by \citet{Cameron1997} were calculated for corresponding models but including nest as a factor with fixed effects instead.

\begin{center}
\begin{table}[htbp]
\caption{\label{Table-Owl}Coefficients of Determination in Analysis of the Owl Data$^{\ast}$}
\centering
\begin{tabular}{c|c c c c c c c c} \hline\hline
Fixed & \multicolumn{2}{c}{\underline{\ \ \ \ \ \ \ \ \ \ $\rho_M^2$\ \ \ \ \ \ \ \ \ \ }} & \multicolumn{2}{c}{\underline{\ \ \ \ \ $\rho_F^2$\ \ \ \ \ }} & & & & \\
Effects & $R_M^2$ & NJS$^a$ &  $R_F^2$ & NJS$^a$ & AIC & BIC & $R_V^2$ & $R_{KL}^2$\\  \hline %
S$\ast$F,S$\ast$T & .277 (.269) & {\bf .684} & .174 & .278 & 5009 & 5040 & .276 (.237) & .226 (.184) \\
S$\ast$F, T & .275 (.269) & {\bf .684} & .174 & .279 & {\bf 5008} & 5034 & .275 (.237) & .226 (.185) \\
S, F, T & .275 (.270) & .682 & .174 & .277 & 5010 & 5032 & .275 (.238) & .225 (.186) \\
F, T & .275 ({\bf .271}) & .682 & .172 & .276 & 5009 & {\bf 5027} & .274 ({\bf .239}) & .225 ({\bf .187}) \\
T & .217 (.214) & .610 & .096 & .139 & 5288 & 5301 & .218 (.181) & .157 (.117) \\
F & .201 (.199) & .639 & .106 & .184 & 5212 & 5226 & .201 (.163) & .175 (.136) \\
\hline\hline %
\multicolumn{9}{l}{$^{\ast}$\footnotesize{With adjusted values in parentheses;} $^a$ \footnotesize{The method by \citet{Nakagawa2017}.}}
\end{tabular}
\end{table}
\end{center}

As shown in Table~\ref{Table-Owl}, the method by \citet{Nakagawa2017} provided much larger estimates of $\rho_M^2$ than $R_M^2$  as well as much larger estimates of $\rho_F^2$ than $R_F^2$ in all models. On the other hand, $R_M^2$ reported very similar values to $R_V^2$ as $R_{KL}^2$ reported slightly smaller values than both of them. However, adjusted versions of $R_M^2$, $R_V^2$, and $R_{KL}^2$ as well as BIC all selected the model with F and T as the best, while AIC selected the model with S$\ast$F and T. The method by \citet{Nakagawa2017} instead have two models including S$\ast$F and T with the highest estimate of $\rho_M^2$ at .684. In general, all coefficients of determination reported larger values for larger models.

\section{Discussion}

Unlike $p$-values that rely on sample sizes to signal the variable significance, the coefficient of determination, a.k.a. $R^2$, measures the proportion of the variation in the response variable explained by a set of predictors. It plays an important role in agricultural, biological, and ecological research, for example, quantifying the heritability of a trait in molecular biology \citep{Visscher2008}. The popularly used mixed-effects models in such studies demand appropriate extension of $R^2$. Our coefficients of determination are well-defined as long as the link and variance functions are specified, such as quasi-models that may rely on quasi-likelihood functions, other than likelihood functions, to obtain parameter estimates. When the first-order derivative of the variance function is constant as for normal and Poisson distributions, our defined $R^2$ for generalized linear mixed models will be reduced to those defined for linear mixed models following the law of total variance. Together with \citet{Zhang2017}, our definitions unify linear models, generalized linear models, linear mixed models, and generalized linear mixed models.

Practice in science may demand further extension or improvement on the proposed measures. Firstly, it may be of interest to measure the utility of a set of predictors beyond others in modeling a response variable. We may define the coefficient of partial determination as shown in \citet{Zhang2017} to measure the proportion of variation in the response variable unexplained by a set of predictors that can be explained by the additional set of predictors. 
Secondly, calculating aforementioned heritability may demand evaluation of the contribution due to certain random effects while controlling the fixed effects of other factors. In this case, we may directly define $\rho_R^2$ based on the law of total variance, though it may be a challenging task, in particular for generalized linear mixed models. Thirdly, our coefficients of determinations are defined for mixed-effects models assuming that observations are randomly sampled within groups and the groups are randomly sampled within the underlying population. However, a same number of observations may be collected within each group although the subpopulation may be of different size, and sometimes it is the other way around. Therefore the coefficients may be modified using different weights on individual variations from different groups.

All of our defined measures for linear mixed model and generalized linear mixed model are implemented in the R package {\tt rsq}, which is publicly available via the Comprehensive R Archive Network (CRAN).
















\end{document}